\documentclass[aps,prx,twocolumn]{revtex4-2}
\usepackage{color}
\usepackage{longtable}
\usepackage{dcolumn}

\newcolumntype{w}[1]{D{.}{.}{#1}}
\usepackage{graphicx}
\usepackage{bm}
\usepackage{bbm}
\usepackage{times}
\usepackage{nicefrac}
\usepackage{amsmath}
\usepackage{amsfonts}
\usepackage{amssymb}
\usepackage{amsthm}

\newcommand{\lbr}{\langle}
\newcommand{\rbr}{\rangle}

\allowdisplaybreaks

\begin{document}
\preprint{Version 1.0}

\title{Relativistic nuclear recoil effects in hyperfine splitting of hydrogenic systems}

\author{Jakub Hevler,  Andrzej Maro\'n,  and Krzysztof Pachucki}

\affiliation{Faculty of Physics, University of Warsaw,
             Pasteura 5, 02-093 Warsaw, Poland}

\date{\today}
\begin{abstract}
Finite nuclear mass $(Z\,\alpha)^2\,m/M\,E_F$  corrections to the hyperfine splitting in hydrogenic systems are calculated using
a combined relativistic heavy particle and nonrelativistic quantum electrodynamics. The obtained results are in disagreement with previous calculations
by Bodwin and Yennie [Phys. Rev. D {\bf 37}, 498  (1988)].
The comparison of improved theoretical predictions with the corresponding measurements in hydrogen reveals $2\,\sigma$ discrepancy,
which may indicate problems with the proton structure corrections.
\end{abstract}

\maketitle

\section{Introduction}
Nuclear recoil (finite nuclear mass) effects in atomic systems are quite difficult to calculate, 
even for simple systems like hydrogen-like ions. 
Although  the  reduced mass accounts for the finite nuclear mass  on the level of  the Schr\"odinger equation, it is not the case for a Dirac equation.
In fact, one has to refer to a full quantum electrodynamic (QED) theory to account for relativistic nuclear recoil effects.
There exist many formulations of two-body bound-state theory in the literature, mostly based on the Bethe-Salpeter approach \cite{BYG}.
In all these formulations, the kernel of the integro-differential equation can only be constructed perturbatively in powers of the fine structure constant $\alpha$. 
Alternative formulation, introduced by Caswell and Lepage \cite{CL},  is the nonrelativistic QED (NRQED), where one constructs perturbatively an effective Lagrangian,
by matching with a full QED theory. This was a very fruitful formulation, which, combined with dimensional regularization
allowed for many important results for simple atomic systems, like the recent one in Ref. \cite{adkins_2023}. 

Another development in bound-state QED was initiated by V. Shabaev \cite{shabaev_85, shabaev_86}, who formulated in 1985 the exact 
nonperturbative formula for the first order in the electron nuclear mass ratio correction to atomic energy levels. 
A few years later this correction was independently rederived  in Ref. \cite{pure_rec}, and since than it has been used for the calculation 
of the Lamb shift in  hydrogen-like ions \cite{artemyev1, artemyev2}. Very recently this original formula has been extended to finite size nuclei \cite{rec_prl},  
to the hyperfine splitting \cite{hfs_formula}, and to higher powers in the $m/M$ mass ratio \cite{hpqed} under the name
heavy particle QED (HPQED).

From the other side, measurement of the Lamb shift and hyperfine splitting  in hydrogen, hydrogen-like ions, and other simple atomic systems
are accurate enough to determine fundamental constants \cite{codata22} and in general to test fundamental interaction theory. 
Particularly accurate is measurement of HFS  of the ground state of the hydrogen atom \cite{Hmaser1, Hmaser2}
$E_\mathrm{hfs}^\mathrm{exp} = 1\,420\,405.751\,768(1)$ kHz.
However, it has the least accuracy of theoretical predictions among  all transitions in the hydrogen atom.
In spite of the fact that the proton size is 5 orders of magnitude smaller than the atomic hydrogen, 
it gives a significant -33 ppm correction to the hydrogen HFS. 
To verify this proton structure effect and to test the standard model of fundamental interactions, 
all other quantum electrodynamics effects shall be rigorously calculated.
Since the original calculation by Bodwin and Yennie \cite{Bodwin:88} of relativistic nuclear recoil effects, 
theoretical predictions have been in a few $\sigma$ disagreement with this measurement. 
The dubious correction was the one related to the proton structure, namely the distribution of the magnetic moment and the proton polarizability.
Despite many further investigations of the proton structure effects, this discrepancy persisted.

In this work we recalculate relativistic nuclear recoil effects of order $(Z\,\alpha)^2\,m/M$ in a system consisting of a lepton of mass $m$
and an arbitrary nucleus of mass $M$, using complementarily  NRQED and HPQED approaches,  
and obtain a result in disagreement with the previous calculations by Bodwin and Yennie in Ref. \cite{Bodwin:88}. 
Consequently, the discrepancy between theoretical predictions and the measured hyperfine splitting in hydrogen is now
slightly decreased and amounts to about $2\,\sigma$. Most probably this remaining discrepancy originates from 
the not well known proton structure. This can be verified by the analogous measurement of  the hyperfine splitting in 
$\mu$H \cite{antognini, vacchi}, where the proton structure effects are much more significant.  
At the same time nuclear recoil effects are also significant, because muon is about 200 times heavier
than the electron. However, our present calculations of recoil effects are accurate enough not only for H, but also for $\mu$H HFS.

\section{Leading order HFS}
Before moving to the main calculations, let us introduce the hyperfine splitting and the basis notation.
In a relativistic framework the interaction between the static point nuclear magnetic moment $\vec \mu$ \begin{align}
V_\mathrm{hfs}(r) =&\  \frac{e}{4\,\pi}\,\vec \mu\cdot\vec\alpha\times\frac{\vec r}{r^3} \label{01}
\end{align}
and an electron leads to the hyperfine splitting of atomic energy levels given by the expectation value
of the magnetic interaction  $E_\mathrm{hfs} =\langle V_\mathrm{hfs}\rangle$ with the Dirac wave functions.
This formula is valid for the infinite nuclear mass only, and it is not obvious
how to account for the finite nuclear mass in this relativistic framework. In fact, later in this work we present
a recently derived \cite{hfs_formula} exact relativistic formula   for the leading recoil correction to HFS  in mass ratio $m/M$. 
Nevertheless, it is more convenient here to perform a nonrelativistic  expansion in $Z\,\alpha$, 
where finite nuclear mass effects can easily be accounted for.
And so, the leading hyperfine splitting for $S$ states is given by
\begin{align}
E^{(4)}_\mathrm{hfs} \equiv &\ E_F =\frac{8}{3}\,\frac{(Z\,\alpha)^4}{n^3}\,\frac{\mu^3}{M\,m}\,\frac{g}{2}\,\langle \vec I\cdot\vec s\rangle\,,
\label{02}
\end{align}
where  $\mu$ is the reduced mass, $g$ is the nuclear $g$-factor defined by
\begin{align}
\vec\mu = \frac{q}{2\,M}\,g\,\vec I\,, \label{03}
\end{align}
and where $q=-Z\,e$ and $e$ is the electron charge.

The next-to-leading hyperfine splitting for an arbitrary nuclear g-factor is conveniently split into three parts
\begin{align}
E^{(5)}_\mathrm{hfs} = &\  E^{(5)}_\mathrm{fns} + E^{(5)}_\mathrm{rec} + E^{(5)}_\mathrm{pol}\,, \label{04}
\end{align}
where the first part is 
\begin{align}
E^{(5)}_\mathrm{fns} =&\ \frac{2\,Z\,\alpha\,m}{\pi^2}\,
\int\frac{d^3k}{k^4}\,\biggl[\frac{G_E(k^2)\,G_M(k^2)}{1+\kappa}-1\biggr]\,E_F \nonumber \\
=&\  -2\,Z\,\alpha\, m\,r_{\rm Z}\,E_F
\,,\label{05}
\end{align}
and where  $G_E$ and $G_M$ are the electric and magnetic form factors of the nucleus, with normalization $G_M(0) = 1+\kappa = g/2$.
It was convenient to write this finite nuclear size correction in terms of the Zemach radius $r_{\rm Z}$ \cite{Zemach} defined by 
\begin{equation}
r_{\rm Z} = \int d^3 r_1 \int d^3 r_2\,\rho_E(r_1)\,\rho_M(r_2)\,|\vec r_1-\vec r_2|, \label{06}
\end{equation}
with $\rho_E$ and $\rho_M$ being the Fourier transforms of $G_E$ and $G_M/(1+\kappa)$.
We point out that different definitions are assumed for $E^{(5)}_\mathrm{fns}$ in the literature. 
Namely, instead of the lepton mass $m$ in Eq. (\ref{05}) one uses the reduced mass $\mu$ of the system \cite{carlson2008}. 
Both definitions have some advantages and disadvantages. Here, we prefer to use Eq. (\ref{05}),
because the remainder $E^{(5)}_\mathrm{rec}$ would not contain any Zemach-like terms.

The second part in Eq. (\ref{04}) $E^{(5)}_\mathrm{rec}$ is the recoil correction obtained from the forward Born amplitude. 
It is considered in the next section, while for the remainder, the nuclear polarizability$E^{(5)}_\mathrm{pol}$, 
we refer to works by others \cite{carlson2008, Ruth2024}.

\section{Leading recoil correction to HFS}
For the derivation of $E^{(5)}_\mathrm{rec}$ we follow Ref. \cite{hfsdirac} and consider again
the total $E^{(5)}_{\rm hfs}$  correction, which can be represented 
by the two-photon exchange forward scattering amplitude. 
In the temporal gauge $A^0=0$ it takes the form  \cite{khriplovich}
\begin{widetext}
\begin{equation}
E^{(5)}_{\rm hfs} = \frac{i}{2}\,\int\frac{d\omega}{2\,\pi}\,
\int\frac{d^3k}{(2\,\pi)^3}\,\frac{1}{(\omega^2-k^2)^2}\,
\biggl(\delta^{ik}-\frac{k^i\,k^k}{\omega^2}\biggr)\,
\biggl(\delta^{jl}-\frac{k^j\,k^l}{\omega^2}\biggr)\,
t^{ji}\,T^{kl}\,\phi^2(0)\,,\label{07}
\end{equation}
where for the point-like spin $1/2$ particle
\begin{align}
t^{ji} =&\ e^2\biggl[
\langle\bar u(t)|\gamma^j\frac{1}{\not\!t\;- \not\!k-m}\,\gamma^i|u(t)\rangle +
\langle\bar u(t)|\gamma^i\frac{1}{\not\!t\;+ \not\!k-m}\,\gamma^j|u(t)\rangle
\biggr] \nonumber \\
=&\ i\,\epsilon^{ijk}\,4\,e^2\,\omega
\,s^k\,\frac{(\omega^2-k^2)}
{(\omega^2-2\,m\,\omega-k^2)\,(\omega^2+2\,m\,\omega-k^2)}\,,\label{08}
\end{align}
and $t$ is the momentum at rest $t=(m,\vec 0)$. Using above expression for the electron $t^{ji}$,
the two-photon exchange correction to the hyperfine splitting becomes
\begin{align}
E^{(5)}_{\rm hfs}  =&\  -2\,e^2\,\phi^2(0)\,\int\frac{d\,\omega}{2\,\pi}\,
\int \frac{d^3k}{(2\,\pi)^3}\,
\frac{\bigl(\omega^2\,\epsilon^{klj}+k^i\,k^k\,\epsilon^{lij}
-k^i\,k^l\,\epsilon^{kij}\bigr)
\,s^j\,T^{kl}}{\omega\,(\omega^2-k^2)\,(\omega^2-k^2-2\,m\,\omega)\,(\omega^2-k^2+2\,m\,\omega)}\,.\label{09}
\end{align} 
$T^{kl}$ is the corresponding virtual Compton scattering amplitude of the nucleus. 
We will assume at the beginning that the nuclear spin $I=1/2$ and consider only the first two terms in $1/M$ expansion,
which are universal and valid for an arbitrary spin nucleus. $T^{kl}$ amplitude for a finite size $I=1/2$ particle is 
\begin{align}
T^{kl} = (Z\,e)^2\biggl[
\langle\bar u(p)|\Gamma^k(k)\frac{1}{\not\!p\;- \not\!k-m}\,\Gamma^l(-k)|u(p)\rangle +
\langle\bar u(p)|\Gamma^l(-k)\frac{1}{\not\!p\;+ \not\!k-m}\,\Gamma^k(k)|u(p)\rangle\biggr]\,, \label{10}
\end{align}
where
\begin{align}
\Gamma^\mu(k) =&\ \gamma^\mu\,F_1 + \frac{i}{2\,M}\,\sigma^{\mu\nu}\,k_\nu\,F_2 \,. \label{11}
\end{align}
$F_1$ and $F_2$  are related to the electric and magnetic form factors by
\begin{align}
G_E(Q^2) =&\ F_1(Q^2) - \frac{Q^2}{4\,M^2}\,F_2(Q^2)\,, \label{12}\\
G_M(Q^2) =&\ F_1(Q^2)+F_2(Q^2)\,, \label{13}
\end{align}
with $Q^2 = k^2-\omega^2$. We  assume the nucleus to be described exclusively by electromagnetic form factors 
and thus neglect the inelastic contribution.
In addition, we neglect also higher order terms in $1/M$  and obtain
\begin{align}
T^{ji} =&\ i\,\epsilon^{ijk}\,4\,(Z\,e)^2\,\omega
\,\frac{\Bigl[\omega^2\,I^k\,(2\,G_M-G_E)\,G_E -k^2\,I^k\,G_M\,G_E - k^k\,(\vec k\cdot\vec I)\,G_M\,(G_M-G_E)\Bigr]}
{(\omega^2-k^2-2\,M\,\omega)\,(\omega^2-k^2 +2\,M\,\omega)}. \label{14}
\end{align}
This formula is valid for an arbitrary spin nucleus, provided the nuclear quadrupole moment is neglected. 
In comparison to Eq.~(44) from Ref. \cite{hfsdirac} we include $\omega^2$ in the denominator of the above equation,
so it is very similar to the point $t^{ji}$ in Eq. (\ref{08}). It does not affect the first two terms in $1/M$ expansion.
The resulting correction to HFS can now be written as
\begin{align}
E^{(5)}_{\rm hfs}  =&\ -\frac{16\,i}{3}\,(Z\,e^2)^2\,\phi^2(0)\,\vec I\,\vec s\,\int\frac{d\omega}{2\,\pi}\,
\int\frac{d^3k}{(2\,\pi)^3}
\frac{T(\omega, k)}{(\omega^2-k^2)^2-4\,M^2\,\omega^2}\,, \label{15}
\end{align}
where 
\begin{align}
T(\omega, k) =&\
\frac{ G_E^2\,\omega^2(2\,k^2 - 3\,\omega^2) + 2\,G_E\,G_M\,(k^4 - 3\,k^2\,\omega^2 + 3\,\omega^4)-G_M^2\,k^2\,\omega^2 }{ (\omega^2-k^2)
[(\omega^2-k^2)^2-4\,m^2\,\omega^2]}\,,
\end{align}
\end{widetext}
The leading  $\sim 1/M$ term is obtained using
\begin{align}
\frac{4\,k^2}{4\,M^2\,\omega^2-k^4}
\approx&\  
-2\,\pi\,i\,\frac{\delta(\omega)}{M}\,, \label{16}
\end{align}
and then $E^{(5)}_{\rm hfs}$ coincides with $E^{(5)}_\mathrm{fns}$ in Eq. (\ref{05}), while the $1/M^2$ term is obtained by neglecting $\omega^2-k^2$
in the denominator containing the large nuclear mass $M$
\begin{align}
E^{(5)}_{\rm rec}  =&\ \frac{16\,i}{3}\,(Z\,e^2)^2\,\phi^2(0)\,\vec I\,\vec s \int_s\frac{d\omega}{2\,\pi}
\int\!\frac{d^3k}{(2\,\pi)^3}
\frac{T(\omega, k)}{4\,M^2\,\omega^2}\,, \label{17}
\end{align}
where $\int_s d\omega$ denotes a symmetric integration around the pole at $\omega=0$.
The same expression can be derived \cite{hfs_formula} from exact formulas presented in the next section.
Because form factors are functions of $ Q^2 = k^2-\omega^2$, we perform at first 
the angular integration in the four-dimensional space and subtract the $1/Q$ term to obtain
\begin{align}
E^{(5)}_{\rm rec}  =&\ -\frac{16}{3}\,(Z\,\alpha)^2\,\frac{\phi^2(0)}{M^2}\, 
\vec I\,\vec s\int\!\frac{dQ}{Q}\,\Big[T-\frac{4\,m}{Q}\,G_M(0)\Big]\,, \label{18}
\end{align}
where  
\begin{align}
T= &\  
 \left(2\,(x-1)-\frac{2}{x}-\frac{1}{x^2}\right)\,G_E\left(Q^2\right)\,G_M\left(Q^2\right) 
 \nonumber \\ &\   
    +\left(\frac{2}{x}+\frac{1}{2\,x^2}\right)
    G^2_E\left(Q^2\right)+\left(\frac{1}{2\,x^2}-\frac{1}{x}\right)
    G^2_M\left(Q^2\right) \,, \label{19}
\end{align}
and $x = \sqrt{1+4\,m^2/Q^2}+1$. 
The integral over $Q$ can be calculated analytically using exponential parametrization of the nuclear form factors
\begin{align}
\rho(Q^2) =&\ \frac{\Lambda^4}{(\Lambda^2+Q^2)^2}\,. \label{20}
\end{align}
Here we present a leading term in $m/\Lambda$ expansion
\begin{align}
E^{(5)}_\mathrm{rec} =&\ - \frac{Z\,\alpha}{\pi}\,\frac{m}{M}\,E_F\,\frac{2}{g}\biggl[
\frac{g^2}{32} - \frac{25\,g}{8} - \frac{9}{8}
\nonumber \\ &\ 
 + \left( \frac{3\, g^2}{16} - \frac{3\, g}{4} - \frac{9}{4} \right) \ln\left(\frac{m}{\Lambda}\right) 
+ \mathcal{O}\Big(\frac{m}{\Lambda}\Big)^2\biggr], \label{21}
\end{align}
which using
\begin{align}
\ln r \equiv &\
\int d^3 r\int d^3r'\,\rho(r)\,\rho(r')\,\ln|\vec r-\vec r{\,'}|
\nonumber \\ =&\ \frac{23}{12} - \gamma - \ln\Lambda\,,
\end{align}
is in apparent agreement with previous calculations \cite{hfsdirac}, except for the fact that the omitted remainder is now of order $\mathcal{O}\big(m/\Lambda\big)^2$.
This means a lack of $\sim m/\Lambda$ terms and equivalently a lack of the Zemach radius $r_Z = 35/(8\,\Lambda)$ in the $m/\Lambda$ expansion in Eq. (\ref{21}).
This is why we use the lepton mass, and not the reduced mass in $E^{(5)}_\mathrm{fns}$ in Eq. (\ref{05}). 
The presence of $\ln\Lambda$ in Eq. (\ref{21}) indicates a large momentum contribution. In fact, for hydrogen, $\Lambda\sim 840$ MeV is not much smaller
than the proton mass; thus, $\Lambda/M$ corrections might be important. Indeed, for hydrogen, one calculates $E^{(5)}_\mathrm{hfs,rec}$  correction
without expansion in a large proton mass, by considering a full two-photon exchange amplitude with measured nuclear form factors \cite{AA2022a}.
It is not the case for $E^{(6)}_\mathrm{rec} $ correction considered in the next section, which is dominated by a momentum exchange of the order of the  lepton mass.

\section{Formulas for relativistic recoil correction to HFS}
The relativistic  $(Z\,\alpha)^2\,E_F$ correction for a point nucleus can be represented as
\begin{align}
E^{(6)}_\mathrm{hfs} =&\ E^{(6)}_\mathrm{Breit} + E^{(6)}_\mathrm{hfsrec}\,, \label{22}
\end{align}
where the nonrecoil part
\begin{align}
E^{(6)}_\mathrm{Breit} =&\ (Z\,\alpha)^6\,\frac{\mu^3}{m\,M}\,g\,\vec I\cdot\vec s\,\frac{4}{3\,n^3}\,\biggl[\frac{11}{6}-\frac{11}{6\,n^2}+\frac{3}{2\,n}\biggr] \label{23}
\end{align}
is the so-called Breit correction \cite{eides:01}, and $E^{(6)}_\mathrm{hfsrec}$ is the nuclear recoil correction in question.
It will be obtained from the exact nonperturbative formula for the relativistic recoil correction to the hyperfine splitting in hydrogen-like ions \cite{hfs_formula}
\begin{align}
E_\mathrm{hfsrec} =&\ E_\mathrm{kin} + E_\mathrm{so} + E_\mathrm{sec}\,, \label{24}
\end{align}
where
\begin{widetext}
\begin{align}
E_\mathrm{sec} = &\  
-i\,\biggl(\frac{4\,\pi\,Z\,\alpha}{2\,M}\,g\biggr)^2\, [I^{ij}\,, I^{kl}]\,  \int_s \frac{d\,\omega}{2\,\pi}\,\frac{1}{\omega}\,
\langle \phi |\alpha^i\,\nabla^j D(\omega)\,G(E_D+\omega)\,\alpha^k\,\nabla^l\,D(\omega)\,| \phi \rangle\,, \label{25}
\\
E_\mathrm{so} = &\  
-\frac{(g-1)}{M^2}\, I^{jk} \int_s \frac{d\,\omega}{2\,\pi}\,\omega\,\langle \phi | D_T^j(\omega)\, G(E_D+\omega)\,D_T^k(\omega) | \phi \rangle\,,  \label{26}
\\
E_\mathrm{kin} = &\  
-\delta_\mathrm{hfs}\frac{i}{M} \int_s \frac{d\omega}{2\,\pi}\, \frac{1}{\omega^2}\lbr \phi | 
 \big[p^j(V_\mathrm{hfs}(\omega)) -\omega\,D_T^j(\omega)\big]\,G(E_D+\omega )\, 
 \big[ p^j(V_\mathrm{hfs}(\omega)) + \omega\,D_T^j(\omega)\big] | \phi \rbr\,, \label{27}
\end{align}
\end{widetext}
where we generalized original formulas to  arbitrary $d=3-2\,\epsilon$ dimensions, and $I^{ij} = \epsilon^{ijk}\,I^k$ at $d=3$.
The subscript $s$ in $\int_s$ denotes a modification of the Feynman contour, where a symmetric integration around the pole at $\omega=0$ is assumed.
The symbol
\begin{align}
G(E) = [E-H_D]^{-1}\,, \label{28}
\end{align}
where
\begin{align}
H_D = \vec\alpha\cdot\vec p + \beta m +V_C(r) \label{29}
\end{align}
stands for the Dirac-Coulomb Green function with the Coulomb potential $V_C$
\begin{align}
V_C(r) =&\ - 4\,\pi\,Z\,\alpha\,\int \frac{d^dk}{(2\pi)^d}\, e^{i\vec{k}\cdot\vec{r}}\, \frac{\rho(\vec k^{\,2})}{\vec k^{\,2}}\,. \label{30}
\end{align}
Moreover,
\begin{align}
D_T^j(\omega,\vec r) =&\ -4\pi Z\alpha \, \alpha^i \, G_{T}^{ij}(\omega,\vec{r})\,, \label{31}
\end{align}
where $G_T$ is the photon propagator in the temporal gauge
\begin{align}
G_T^{ij}(\omega, \vec r) =&\ \int \frac{d^dk}{(2\pi)^d}\, e^{i\vec{k}\cdot\vec{r}}\,
\frac{\rho({\vec k}^{\,2}-\omega^2)}{\omega^2-{\vec k}^2}\,\biggl(\delta^{ij}-\frac{k^i\,k^j}{\omega^2}\biggr)\,. \label{33}
\end{align}
The photon propagator, as well as the Coulomb potential, includes the nuclear form factor,
which we assume for simplicity to be the same for the electric and the magnetic one.
The symbol $\delta_\mathrm{hfs}$ in Eq. (\ref{27}) denotes the first-order correction in the hyperfine interaction.
This interaction is explicit as a $V_\mathrm{hfs}$, and implicit in the wave function $\phi$, 
the corresponding Dirac energy $E_D$, and in the fermion propagator $G$ by adding $V_\mathrm{hfs}$ to the Dirac Hamiltonian.
Moreover, Eq. (\ref{27}) includes the frequency-dependent  hyperfine potential
\begin{align}
V_\mathrm{hfs}(\omega,\vec r) =&\  
-\frac{Z\,\alpha}{2\,M}\,g\,I^{ij}\,\alpha^i\,\nabla^j 4\,\pi\,D(\omega, r)\,, \label{35}
\end{align}
where
\begin{align}
D(\omega,r) =&\ \int \frac{d^dk}{(2\pi)^d}\, e^{i\vec{k}\cdot\vec{r}}\,\frac{\rho({\vec k}^{\,2}-\omega^2)}{\omega^2-{\vec k}^2}\,, \label{36}
\end{align}
and $V_\mathrm{hfs}(\vec r) = V_\mathrm{hfs}(0,\vec r)$. 

We now pass to the calculation of $(Z\,\alpha)^2\,m/M\,E_F$ correction to the hyperfine splitting
in hydrogenic systems.  We note that $\int_s \omega^{-n}$ vanishes for an integer $n>1$. 
Therefore, due to $\pm\omega$ symmetrization, we can subtract from the integrand all $\omega=0$ singularities.
Moreover, for $(Z\,\alpha)^2\,m/M E_F$ correction we can set $\rho=1$, which is not necessarily obvious.
We will use dimensional regularization, perform Wick rotation $\omega\rightarrow i\,\omega$, 
and split the recoil correction into the low- and high-energy parts
\begin{align}
E^{(6)}_\mathrm{hfsrec} =&\ E_L + E_H\,. \label{37}
\end{align}
In the low-energy part $\vec k\sim m\,\alpha$, and in the high-energy part $\vec k\sim m$.
In the dimensional regularization  the low- and high-energy parts can be calculated separately
using a different $\omega$-integration contour and a different expansion adjusted to the particular energy scale. 

\section{High-energy part: HPQED}
The high-energy part is obtained by the hard three-photon exchange approximation of HPQED formulas in  Eqs. (\ref{25},\ref{26},\ref{27}), namely
\begin{widetext}
\begin{align}
E_\mathrm{secH} = &\ 
-i\,\Bigl(\frac{g}{2\,M}\Bigr)^2\,(Z\,\alpha)^3\,\phi^2(0)\,[I^{ij}\, I^{kl}] \int_s \frac{d\,\omega}{2\,\pi}\,\frac{1}{\omega}
\int\frac{d^dk_1}{(2\,\pi)^d} \int\frac{d^dk_2}{(2\,\pi)^d}\,\frac{(4\,\pi)^3\,k_2^j\,k_1^l}{k_1^2\,k_2^2\,(k_1-k_2)^2}\,X^{ik0}\,, \label{72}
\\
E_\mathrm{soH} = &\  
\frac{(g-1)}{M^2}\,(Z\,\alpha)^3\,\phi^2(0)\,I^{jk}\!\! \int_s \frac{d\,\omega}{2\,\pi}\,\omega\!
\int \frac{d^3k_1}{(2\,\pi)^3} \int \frac{d^3k_2}{(2\,\pi)^3}
\frac{4\,\pi}{k_2^2}\,\frac{4\,\pi}{k_{12}^2}\,\frac{4\,\pi}{k_1^2}\, 
\biggl(\delta^{jb}-\frac{k_1^j\,k_1^b}{\omega^2}\biggr)\, \biggl(\delta^{ka}-\frac{k_2^k\,k_2^a}{\omega^2}\biggr)X^{ab0}\,, \label{73}
\end{align}
where the momentum square $k^2$ is in $D=d+1$ dimensions.
The high-energy contribution $E_\mathrm{kinH}$ is split into two parts
$E_\mathrm{kinH} = E_\mathrm{kinHA} + E_\mathrm{kinHB}$.
The first part comes from explicit presence of $V_\mathrm{hfs}(\omega)$ in Eq. (\ref{27})
\begin{align}
E_\mathrm{kinHA} = &\  
-\frac{g}{2\,M^2}(Z\,\alpha)^3 \phi^2(0)\,I^{ij}\! \int_s \frac{d\omega}{2\,\pi}\,\frac{1}{\omega}
\int \! \frac{d^d k_1}{(2\,\pi)^d} \int \! \frac{d^d k_2}{(2\,\pi)^d}
\biggl(\delta^{kl}-\frac{k_2^k\,k_2^l}{\omega^2}\biggr) k_1^k\,k_1^j\,\frac{(4\,\pi)^3}{k_1^2\,k_2^2\,k_{12}^2}
\big[X^{li0}(k_2, k_1) -X^{il0}(k_1, k_2)\big],
\end{align}
and the second part comes from implicit dependence on $V_\mathrm{hfs}$ in $H_D, E_D$, and $\phi$ in Eq. (\ref{27})
\begin{align}
E_\mathrm{kinHB} =&\ 
-\frac{g}{2\,M^2} (Z\,\alpha)^3 \phi^2(0)\,I^{ab}\! \int_s \frac{d\omega}{2\,\pi}
\int\frac{d^d k_1}{(2\,\pi)^d} \int\frac{d^d k_2}{(2\,\pi)^d}
\biggl(\delta^{ik}-\frac{k_2^i\,k_2^k}{\omega^2}\biggr)
\biggl(\delta^{jk}-\frac{k_1^j\,k_1^k}{\omega^2}\biggr)
\frac{(4\,\pi)^3}{k_1^2\,k_2^2\,k_{12}^2}\,(k_1-k_2)^b\,X^{ija},
\end{align}
where $X^{ij\mu}$ is
\begin{align}
X^{ij\mu}(k_2,k_1)=&\ \frac{\gamma^0+I}{2}\,\biggl[
 \gamma^i\,\frac{1}{\not\!k_2+\not\!t-m}\,\gamma^\mu\,\frac{1}{\not\!k_1+\not\!t-m}\,\gamma^j 
 + \gamma^\mu\,\frac{1}{\not\!k_1 -\not\!k_2+\not\!t-m}\,\gamma^i\,\frac{1}{\not\!k_1+\not\!t-m}\,\gamma^j
 \nonumber \\ &\ 
 +  \gamma^i\,\frac{1}{\not\!k_2+\not\!t-m}\,\gamma^j\,\frac{1}{\not\!k_2 - \not\! k_1+\not\!t-m}\,\gamma^\mu 
 \biggr]\frac{\gamma^0+I}{2}\,. \label{76}
  \end{align}
After performing spin algebra, described in Appendix  \ref{AppA}, $E_H$ becomes an integral over scalar function of $\vec k_1, \vec k_2$, and $\omega$.
Using formulas from Appendix \ref{AppD}, we integrate at first over $k_i$ and next over $\omega$, and divide by factor $\eta$ in Eq. (\ref{C4})
from the dimensional regularization, to  obtain
\begin{align}
E_\mathrm{secH} =&\
\frac{(Z\,\alpha)^6}{n^3}\,\frac{m^3}{M^2}\,g^2\,I^{ij}\, \sigma^{ij}
\frac{1}{96}\,\bigg(-\frac{50}{3} - \frac{7}{\epsilon} -28\,\ln(Z\,\alpha) + 32\,\ln 2\bigg)\,, \label{78}
\\
E_\mathrm{soH} = &\  
\frac{(Z\,\alpha)^6}{n^3}\,\frac{m^3}{M^2}\,(g-1)\,I^{ij}\,\sigma^{ij}\,
\frac{1}{3}\,\bigg(-\frac{83}{12} + \frac{7}{8\,\epsilon} + \frac{7}{2}\,\ln(Z\,\alpha) + 8\,\ln 2\bigg)\,, \label{79}
\\
E_\mathrm{kinHA} = &\ 
\frac{(Z\,\alpha)^6}{n^3}\,\frac{m^3}{M^2}\,g\,I^{ij}\,\sigma^{ij}\,\biggl(\frac{65}{36} -\frac{1}{6\,\epsilon} -4\,\ln2 -\frac{2}{3}\,\ln(Z\,\alpha)\biggr)\,, \label{80}
\\
E_\mathrm{kinHB} =&\ 
-\frac{(Z\,\alpha)^6}{n^3}\,\frac{m^3}{M^2}\,g\,I^{ij}\,\sigma^{ij}\,\frac{7}{12}\,, \label{81}
\end{align}
\end{widetext}
where $\sigma^{ij} = \epsilon^{ijk}\,\sigma^k$ and $I^{ij} = \epsilon^{ijk}\,I^k$ in $d=3$,
which concludes the calculation of the high-energy part.

\section{Low-energy part: NRQED}
For the calculation of the low-energy part, instead of the direct use of the above formulas, we will use an equivalent  NRQED formalism because it is much simpler.
Corrections to energy levels can be written as an expectation value of the effective Hamiltonian, namely
\begin{align}
E_L = \langle H^{(6)}\rangle + \bigg\langle H^{(4)}\,\frac{1}{(E-H)'}\,H^{(4)}\bigg\rangle, \label{38}
\end{align}
where $H^{(4)}$ is the so-called Breit Hamiltonian. For the two-body system, neglecting spin-orbit terms which vanish for S-states, it takes a form
\begin{align}
 H^{(4)}  = &\ H^{(4)}_\mathrm{ns} + H^{(4)}_\mathrm{hfs}\,, \label{39}
 \end{align}
 where
\begin{align}
 H^{(4)}_\mathrm{ns} =&\
 \sum_{a=1,2}\biggl\{ -\frac{p_a^4}{8\,m_a^3} +\frac{Z\alpha}{2\,m_a^2}\,\pi\,\delta^d(r)\biggr\}
 \nonumber \\ &\
+\frac{Z\alpha}{2\,m_1\,m_2}\,p_1^i\,
\biggl[\frac{\delta^{ij}}{r}+\frac{r^i\,r^j}{r^3}
\biggr]_\epsilon\,p_2^j\,, \label{40}
\\ 
H^{(4)}_\mathrm{hfs} =&\
\frac{Z\,\alpha\,g}{m\,M}\, I^{ij}\,\sigma^{ij}\,\frac{\pi}{d}\,\delta^d(r)
\nonumber \\ &\
+\frac{Z\,\alpha\,g}{4\,m\,M}\,I^{ij}\,\sigma^{ik}\,\biggl[\frac{\delta^{jk}}{r^3} - 3\,\frac{r^j\,r^k}{r^5}\biggr]_\mathrm{\epsilon}\,, \label{41}
\end{align}
where $\vec r = \vec r_1 - \vec r_2$, $e_1 = e$, $e_2 = -Z\,e$, $\sigma^{ij} = \sigma^{ij}_1$,  $I^{ij} = \sigma^{ij}_2/2$,
and the subscript $\epsilon$ stands for the dimensionally regularized form of an operator written explicitly in $d=3$, 
which is described in detail in Appendix \ref{AppB}.

Derivation of $H^{(6)}$ is presented in the following, and we heavily rely on Refs. \cite{hesinglet, twobodyp}. 
The NRQED Hamiltonian for an electron (particle 1) is
\begin{align}
H =&\ e\,A^0 + \frac{\vec \pi^{\,2}}{2\,m} + \frac{e}{4\,m}\,\sigma^{ij}\,B^{ij} -\frac{\vec \pi^{\,4}}{8\,m^3} -\frac{e}{8\,m^2}\,\vec\nabla\vec E 
\nonumber \\ &\ 
-\frac{e}{8\,m^2}\,\sigma^{ij}\,\{E^i\,,\,\pi^j\}
+\frac{e}{16\,m^3}\,\sigma^{ij}\,\{B^{ij}\,,\,\vec p^{\,2}\}\,, \label{42}
\end{align}
and for the nucleus (particle 2)  we neglect $1/M^3$ terms but assume an arbitrary $g$
\begin{align}
H =&\ e\,A^0 + \frac{\vec \pi^{\,2}}{2\,m} + \frac{e\,g}{8\,m}\,\sigma^{ij}\,B^{ij} -\frac{e\,(g-1)}{8\,m^2}\,\sigma^{ij}\,\{E^i\,,\,\pi^j\}, \label{43}
\end{align}
where  $B^{ij} = \epsilon^{ijk}\,B^k$ in $d=3$ case.
Using these NRQED Hamiltonians, one obtains an effective 
$H^{(6)}_\mathrm{hfs}$ interaction Hamiltonian, which is a sum of 5 terms
\begin{align}
H^{(6)}_\mathrm{hfs} =&\ \sum_{i=1,5} \delta H_i\, \label{44}
\end{align}
calculated below. Let us define static fields ${\cal A}^0$, $\vec {\cal A}$, and $\vec {\cal E}$ by
\begin{align}
e_1\,{\cal A}_1^0 =&\ e_2\,{\cal A}^0_2 = -\frac{Z\,\alpha}{r_\epsilon}\,, \label{45}\\
e_1 {\cal{A}}^{i}_{1} =&  - \frac{Z\,\alpha\,g_2}{4\,m_2}\,\sigma_2^{ki}\,\biggl(\frac{r^k}{r^3}\biggr)_\epsilon\,, \label{46}
\\		
e_2 {\cal{A}}^{i}_{2} =& \frac{Z\,\alpha\,g_1}{4\,m_1}\,\sigma_1^{ki}\,\biggl(\frac{r^i}{r^3}\biggr)_\epsilon\,, \label{47} \\
e_{1}\,\vec{\cal{E}}_1 =& - Z\,\alpha\, \biggl(\frac{\vec{r}}{r^3}\biggr)_\epsilon,\,\,
e_{2}\,\vec{\cal{E}}_2 =  Z\,\alpha\, \biggl(\frac{\vec{r}}{r^3}\biggr)_\epsilon \,, \label{48}
\end{align}
with $g_1 = 2$. We now examine the individual contributions $\delta H_i$.
$\delta H_1$ is a correction to the Coulomb interaction when both vertices are
\begin{align}
- \frac{e\,(g-1)}{8\, m^2}\,\sigma^{ij}\,\{E^i\,,\pi^j\}\,. \label{49}
\end{align}
It can be evaluated in the nonretardation approximation, with the result
\begin{align}
\delta H_{1} =&\ -Z\,\alpha\,\frac{(g_1-1)}{m_1^2}\,\frac{(g_2-1)}{m_2^2}
\int \frac{d^d k}{(2\,\pi)^d}\,\frac{4\,\pi}{k^2}\,\frac{1}{64}\,
\nonumber \\ &\times
\bigl(-2\,i\,\sigma_1^{ij}\,k^i\,p_1^j\bigr)\,
e^{i\,\vec k\cdot\vec r}\,
\bigl(2\,i\,\sigma_2^{kl}\,k^k\,p_2^l\bigr)
\nonumber \\ =&\ 
\frac{(g_1-1)}{m_1^2}\,\frac{(g_2-1)}{m_2^2}\,
\frac{Z\,\alpha\,\sigma_1\,\sigma_2}{16\,d(d-1)}\,p^i\,4\,\pi\,\bigl[\delta^{ij}_\perp (r)\bigr]_\epsilon\,p^j. \label{50}
\end{align}
$\delta H_2$ is the relativistic correction to the transverse photon exchange. The second particle is coupled to $\vec A$ by the nonrelativistic term
\begin{align}
- \frac{e\,g}{8\,m}\, \sigma^{ij}\,B^{ij}\,, \label{51}
\end{align}
and the first one by the relativistic correction
\begin{align}
	     \frac{e}{16\,m^3}\,\bigl\{\vec p^{\,2}, \sigma^{ij}\,B^{ij}\bigr\}\,. \label{52}
\end{align}
It is sufficient to calculate it in the nonretardation approximation, which yields
\begin{align}
	\delta H_2 &\ = \frac{1}{8\,m_1^{3}}\{p^{2} ,\, \sigma_1^{ij}\,\nabla_1^i\,e_{1} {\cal{A}}_1^j\}
	\nonumber \\ =&\
	 -\frac{Z\,\alpha\,g_2\,\sigma_1\sigma_2}{16\,d\,m_1^{3}\,m_2}\,p^{2}\,4\,\pi\,\delta^d(r)\,, \label{53}		 
\end{align}
where $\sigma_1\,\sigma_2 = \sigma_1^{ij}\,\sigma_2^{ij}$.
$\delta H_3$ is a seagull-like term that comes from
\begin{align}
 \frac{e^2\,(g-1)}{4\, m^2}\,\sigma^{ij}\,E^i\,A^j\,. \label{54}
\end{align}
Once more the nonretardation approximation can be used, yielding
\begin{align}
	\delta H_3 =& \sum_{a} \, \frac{e_{a}^2\,(g_a-1)}{4\, m_{a}^2} \, \sigma_{a}^{ij}\,{\cal{E}}_a^i\,{\cal{A}}_a^j
	\nonumber \\ =&\
	\frac{\sigma_1\,\sigma_2}{d}\,\frac{(Z\,\alpha)^2}{8}\,\frac{1}{r^4_\epsilon} \,\biggl(\frac{g_2}{2\,m_1^2\,m_2} + \frac{g_2-1}{m_2^2\,m_1}\biggr)\,.
	\label{55}
	\end{align}
$\delta H_4$ is a retardation correction in a single transverse photon exchange, where one vertex is nonrelativistic, 
\begin{align}
- \frac{e\,g}{8\,m}\, \sigma^{ij}\,B^{ij}\,, \label{56}
\end{align}
 and the second one is
\begin{align}
 - \frac{e\,(g-1)}{8\, m^2}\,\sigma^{ij}\,\{E^i\,,\,p^j\}\,. \label{57}
\end{align}
The result is
\begin{align}
\delta H_4 =&\ \sum_{a} \frac{e_{a}^2\,(g_a-1)}{4\, m_{a}^2}\, \sigma_a^{ij}\,{\cal E}_a^i\,{\cal A}_a^j
\nonumber \\ &\
	- \frac{i\,e_{a}\,(g_a-1)}{16\,m_{a}^3}\, \big[ \sigma_a^{ij}\, \{p_a^i\,,\,{\cal A}_a^j\}\,,\, \vec p_{a}^{\,2} \, \big]
\nonumber \\ =&\
\frac{\sigma_1\,\sigma_2}{8\,d}\,\biggl[
\frac{g_2}{2\,m_1^2\,m_2} + \frac{g_2-1}{m_2^2\,m_1}
-\frac{g_2\,\mu}{2\,m_1^3\,m_2}\biggr]\,\frac{(Z\,\alpha)^2}{r^4_\epsilon}. \label{58}
\end{align}
$\delta H_5$ is a retardation correction to the single transverse exchange 
where both vertices are
\begin{align}
- \frac{e\,g}{8\,m}\, \sigma^{ij}\,B^{ij}\,. \label{59}
\end{align}	
The result is
\begin{align}
\delta H_5 =& -\frac{Z\,\alpha\,g_2}{32\,m_1^2\,m_2^2}\,\frac{\sigma_1\,\sigma_2}{d}\,
\biggl[p^2,\biggl[p^2,\biggl[\frac{1}{r}\biggr]_\epsilon\biggr]\biggr]
\nonumber \\ =&\
-\frac{(Z\,\alpha)^2\,\mu\,g_2}{8\,m_1^2\,m_2^2}\,\frac{\sigma_1\,\sigma_2}{d}\,\frac{1}{r^4_\epsilon}\,, \label{60}
\end{align}
where
\begin{align}
\biggl\langle\bigg[p^2\,,\,\bigg[p^2\,,\,\frac{1}{r_\epsilon}\bigg]\bigg]\biggr\rangle =&\ 4\,Z\,\alpha\,\mu\,\Big\langle\frac{1}{r^4_\epsilon}\Big\rangle\,. \label{61}
\end{align}
This concludes our derivation of all spin-spin effective operators to order $\alpha^6$.
The total effective Hamiltonian $H^{(6)}_\mathrm{hfs}$ after transformation to atomic units
and dividing by a factor $\eta$ from Eq. (\ref{C4}) is
\begin{align}
H^{(6)}_\mathrm{hfs} =&\ H^{(6)}_\mathrm{so} + H^{(6)}_\mathrm{kin} + O(M^{-3})\,, \label{62}
\end{align}
where
\begin{align}
H^{(6)}_\mathrm{so} =&\
(Z\,\alpha)^6\,\frac{I^{ij}\,\sigma^{ij}}{2\,d}\,(g_2-1)\,\frac{\mu^3}{M^2}
\nonumber \\ &\times
\biggl[\frac{1}{(d-1)}\,\pi\,\bigl[\delta^{ij}_\perp (r)\bigr]_\epsilon\,p^i\,p^j +\frac{1}{r^4_\epsilon}\biggr]\,, \label{63}
\\
H^{(6)}_\mathrm{kin} =&\
(Z\,\alpha)^6\,\frac{I^{ij}\,\sigma^{ij}}{8\,d}\,g_2\,\frac{\mu^5}{M\,m^3}\biggl[ -p^{2}\,4\,\pi\,\delta^d(r)+\frac{1}{r^4_\epsilon}\biggr]\,. \label{64}
\end{align}

Let us now obtain individual low-energy parts for $E_\mathrm{sec}$,  $E_\mathrm{so}$, and  $E_\mathrm{kin}$.
$H^{(6)}$ does not contain $g^2$, thus $E_\mathrm{secL}$ comes only from second-order terms
\begin {align}
E_\mathrm{secL} =&\ \langle \phi | H^{(4)}_\mathrm{hfs}\,\frac{1}{(E-H)'}\,H^{(4)}_\mathrm{hfs}| \phi \rangle_\mathrm{hfs}\,. \label{65}
\end{align}
It consists of two parts $E_\mathrm{secL} = E_\mathrm{secLA} + E_\mathrm{secLB}$.  
The first part $E_\mathrm{secLA}$ comes from the second-order contact  interactions.
Using matrix elements from Appendix \ref{AppB}, it is
\begin{align}
E_\mathrm{secLA} =&\ 
-(Z\,\alpha)^6\,\frac{m^3}{M^2}\,g^2\,I^{ij}\, \sigma^{ij}\,\frac{2}{9\,n^3}
\nonumber \\ & \times
\bigg(- \frac13  - \frac{1}{n} + \gamma + \Psi(n) - \ln\frac{n}{2}  -\frac{1}{4\,\epsilon}\bigg)\,. \label{66}
\end{align}
Similarly, the second part $E_\mathrm{secLB}$ comes from the second-order tensor interaction
\begin{align}
E_\mathrm{secLB} =&\ 
-(Z\,\alpha)^6\,\frac{m^3}{M^2}\,g^2\,I^{ij}\, \sigma^{ij}\,\frac{5}{72\,n^3}
\nonumber \\ &\hspace*{-9ex}\times
\bigg(-\frac{19}{20} + \frac{1}{2\,n} + \frac{3}{20\,n^2} + \gamma + \Psi(n) - \ln\frac{n}{2} -\frac{1}{4\,\epsilon}\bigg)\,. \label{67}
\end{align}

The spin-orbit contribution $E_\mathrm{soL}$ does not contain second-order terms, 
only the first-order ones $H^{(6)}_\mathrm{so}$. Using matrix elements from Appendix \ref{AppB}, we obtain 
\begin{align}
E_\mathrm{soL} =&\  \langle H^{(6)}_\mathrm{so}\rangle
\nonumber \\ =&\
(Z\,\alpha)^6\,\frac{m^3}{M^2}\,I^{ij}\,\sigma^{ij}\,(g-1)\,\frac{1}{6\,n^3}\,\biggl[
-\frac{67}{12}  + \frac{7}{2\,n} 
\nonumber \\ &\
+ \frac{11}{12\,n^2} + 7\,\bigg(\ln 2 - \ln n + \gamma + \psi(n) - \frac{1}{4\,\epsilon} \bigg)\biggr]. \label{68}
\end{align}

The kinetic part  $E_\mathrm{kinL}$ contains the first- and the second-order terms, thus
\begin{align}
E_\mathrm{kinL} =&\ \langle H^{(6)}_\mathrm{kin}\rangle + 
2\,\bigg\langle H^{(4)}_\mathrm{ns}\,\frac{1}{(E-H)'}\,H^{(4)}_\mathrm{hfs}\bigg\rangle 
\nonumber \\ =&\
E_\mathrm{Breit} + E'_\mathrm{kinL}\,, \label{69}
\end{align}
where
\begin{align}
E_\mathrm{Breit} =&\
(Z\,\alpha)^6\,\frac{\mu^3}{m\,M}\,\frac{g}{2}\,\vec I\cdot\vec s\,\frac{8}{3\,n^3}\,\biggl[\frac{11}{6}-\frac{11}{6\,n^2}+\frac{3}{2\,n}\biggr]\,, \label{70}
\\
E'_\mathrm{kinL} =&\
\frac{(Z\,\alpha)^6}{n^3}\,\frac{m^3}{M^2}\,g\,I^{ij}\,\sigma^{ij}\,\biggl[
\frac{8}{9} - \frac{1}{3\,n}+ \frac{7}{18\,n^2} 
\nonumber \\ &\
-\frac{2}{3}\,\biggl( \ln 2 - \ln n + \gamma + \psi(n) - \frac{1}{4\,\epsilon}\biggr)
\biggr]\,, \label{71}
\end{align}
which completes the calculation of the low-energy part. 

\section{Results for relativistic recoil HFS}
Final formulas for all contributions $E^{(6)}_\mathrm{hfsrec}$ are obtained from
\begin{align}
E^{(6)}_\mathrm{hfsrec} =&\ E^{(6)}_\mathrm{sec} + E^{(6)}_\mathrm{so}  + E^{(6)}_\mathrm{kin}\,, \label{82}
\end{align}
where
\begin{align}
E^{(6)}_\mathrm{sec} =&\ E_\mathrm{secLA}  +  E_\mathrm{secLB} +  E_\mathrm{secH}\,, \label{83} \\
E^{(6)}_\mathrm{so} =&\ E_\mathrm{soL} +  E_\mathrm{soH}\,, \label{84} \\
E^{(6)}_\mathrm{kin} =&\ E'_\mathrm{kinL} + E_\mathrm{kinHA} + E_\mathrm{kinHB}\,, \label{85}
\end{align}
and take the form
\begin{widetext}
\begin{align}
E^{(6)}_\mathrm{sec} =&\ 
(Z\,\alpha)^6\,\frac{m^3}{M^2}\,g^2\,\vec I\cdot\vec s\,\frac{2}{3\,n^3}\biggl[
\frac{1}{4}\,\ln 2 +\frac{31}{36} 
+ \biggl( \frac{9}{8\,n} - \frac{1}{16\,n^2}  - \frac{17}{16} \biggr)
- \frac{7}{4}\,\Bigl(\gamma + \Psi(n) - \ln n + \ln(Z\,\alpha)\Big) \biggr]\,, \label{86}
\\
E^{(6)}_\mathrm{so} =&\ 
(Z\,\alpha)^6\,\frac{m^3}{M^2}\,(g-1)\,\vec I\cdot\vec s\,\frac{8}{3\,n^3}\,\biggl[
\frac{23}{4}\,\ln 2 - \frac{15}{4} + \biggl(\frac{7}{8\,n} + \frac{11}{48\,n^2} -\frac{53}{48}\biggr) + \frac{7}{4}\,\Bigl(\gamma + \Psi(n) -\ln n + \ln(Z\,\alpha)\Bigr) \biggr]\,,
\label{87}
\\
E^{(6)}_\mathrm{kin} =&\ 
(Z\,\alpha)^6\,\frac{m^3}{M^2}\,g\, \vec I\cdot\vec s\,\frac{4}{3\,n^3}\biggl[
-14\,\ln 2 + \frac{13}{2}  + \biggl( -\frac{1}{n} +\frac{7}{6\,n^2}-\frac{1}{6}\biggr)  - 2\,\Big(\gamma +\Psi(n) -\ln n +\ln(Z\,\alpha)\Big) \biggr]\,.
\label{88}
\end{align}
The sum for $g=2$ and $I=1/2$ is
\begin{align}
E^{(6)}_\mathrm{hfs} =&\
(Z\,\alpha)^6\,\frac{\mu^3}{m_1\,m_2}\,\frac{8}{3\,n^3}\,\biggl[\frac{11}{6}-\frac{11}{6\,n^2}+\frac{3}{2\,n}\biggr]
\nonumber \\ &\ 
+ (Z\,\alpha)^6\,\frac{\mu^5}{m_1^2\,m_2^2}\,\frac{8}{3\,n^3}\,\biggl\{
- 2\,\Big(\gamma +\Psi(n) -\ln n +\ln(Z\,\alpha)\Big) + \biggl(\frac{4}{3\,n^2} +\frac{1}{n} - \frac{7}{3}\biggr) -8\,\ln 2 +\frac{65}{18}
\biggr\}\,,  \label{89}
\end{align}
in agreement with previous calculations in Ref. \cite{pachucki:97}  and also for $n=1$ with \cite{BYG} .
Our result for the 1S state is
\begin{align}
E^{(6)}_\mathrm{rec}(1S) =&\ 
\bigg[\frac{65}{18} + \frac{13}{18}\,\kappa + \frac{31}{36}\,\kappa^2
- \bigg(8 + 2\,\kappa - \frac{1}{4}\,\kappa^2\bigg)\,\ln2 
- \bigg(2 + 2\,\kappa + \frac{7}{4}\,\kappa^2\bigg)\,\ln(Z\,\alpha)\bigg]\,(Z\,\alpha)^2\,\frac{m}{M}\,\frac{E_F}{1+\kappa}\,, \label{90}
\end{align}
 in disagreement with the previous result of Bodwin and Yennie \cite{Bodwin:88}
\begin{align}
E^{(6)}_\mathrm{BY}(1S) =&\ 
\bigg[\frac{65}{18} + \frac{43}{72}\,\kappa + \frac{31}{36}\,\kappa^2
- \bigg(8 + 5\,\kappa + \frac{11}{4}\,\kappa^2\bigg)\,\ln2 
- \bigg(2 + 2\,\kappa + \frac{7}{4}\,\kappa^2\bigg)\,\ln(Z\,\alpha)\bigg]\,(Z\,\alpha)^2\,\frac{m}{M}\,\frac{E_F}{1+\kappa} .\label{91}
\end{align}
\end{widetext}
We were not able to identify errors in Bodwin and Yennie \cite{Bodwin:88} calculations, but from our side we have checked
our result for $E_\mathrm{sec}$ by direct numerical calculation of Eq. (\ref{25}) for the ground state $(n=1)$, 
and this will be described in detail in a subsequent work \cite{Hevler}.

\section{Improved theory of hyperfine splitting in hydrogenic systems}
The complete hyperfine splitting is conveniently represented as
\begin{align}
E_\mathrm{hfs} =&\ E_F\,(1 + \delta)\,,  \label{116}
\end{align}
where $\delta$ represents corrections
due to relativistic, QED, and nuclear effects. 
Expansion of $\delta$  in powers of the fine-structure constant $\alpha$ is,
\begin{align} \label{117}
\delta =&\  a_e + \delta^{(2)}+ \delta^{(3)} + \delta^{(4)} + \delta^{(1)}_\mathrm{fns}  + \delta^{(1)}_\mathrm{rec} + \delta^{(1)}_\mathrm{pol}
\nonumber \\ &\
+ \delta^{(2+)}_\mathrm{rel,fns} + \delta^{(2)}_\mathrm{rad,fns}  + \delta^{(2)}_\mathrm{rel,rec} + \delta^{(2)}_\mathrm{rad,rec}  
\nonumber \\ &\
+ \delta^{(2)}_\mathrm{\mu vp} + \delta^{(2)}_\mathrm{hvp}  + \delta_\mathrm{weak}\,.
\end{align}
Let us now explain various corrections in accordance with previous reviews on this topic 
\cite{eides:01,SGK2005,volotka}. 

$a_e $ is the magnetic moment anomaly of a free electron, which is known to an accuracy of about $10^{-10}$ \cite{codata22}.
$\delta^{(i)}$ are QED corrections of order $\alpha^i$, which are well known except for $\delta^{(4)}$,
\begin{align}
\delta^{(2)}&\  =\frac{3}{2}\,(Z\,\alpha)^2 + \alpha\,(Z\,\alpha) \Bigl(\ln(2)-\frac{5}{2}\Bigr)\,, \label{118} \\
\delta^{(3)}&\  = \frac{\alpha\,(Z\,\alpha)^2}{\pi}\!
\Big[\!-\frac{8}{3}\ln(Z\,\alpha)\Bigl(\ln(Z\,\alpha)-\ln(4)+\frac{281}{480}\Bigr)
\nonumber \\ &\
+ 17.122\,338\,751\,3-\frac{8}{15}\,\ln(2)+\frac{34}{225}\Bigr]
\nonumber \\ &\
+ \frac{\alpha^2\,(Z\,\alpha)}{\pi}\,0.770\,99(2) \,,  \label{119} \\
\delta^{(4)} &\  =\frac{17}{8}\,(Z\,\alpha)^4 +\alpha\,(Z\,\alpha)^3\,\Big[\Big(\frac{547}{48}-5\,\ln(2) \Big)\,\ln(Z\,\alpha)
\nonumber \\ &\
 -4.402\,5(13) +\frac{13}{24}\,\ln 2 + \frac{539}{288} \Bigr]
\nonumber
\\ &\
-\frac{\alpha^2\,(Z\,\alpha)^2}{\pi^2}\,\Bigl[\frac{4}{3}\,\ln^2(Z\,\alpha)+1.278\,\ln(Z\,\alpha) +10.0(2.5)\Bigr]
  \nonumber \\ & \
\pm  \frac{\alpha^3\,(Z\,\alpha)}{\pi^2}\,.  \label{120}
\end{align}
Most of the results summarized by Eqs.~(\ref{08})-(\ref{10}) can be found in Refs.~\cite{eides:01,codata22}.
We mention that the $\alpha(Z\,\alpha)^2$ part of $\delta^{(3)}$ contains the improved numerical value 
for the constant term from Ref. \cite{heplus}, and the $\alpha(Z\,\alpha)^3$ part of $\delta^{(4)}$ includes higher orders 
in $Z\,\alpha$ for $Z=1$ from Ref.~\cite{yerokhin:08:prl}.
The last term in $\delta^{(4)}$ represents the estimate of the unknown three-loop QED binding correction. 
We should mention that these QED corrections partially include nuclear recoil (or finite nuclear mass) effects  through $\phi^2(0)$ in $E_F$.
Further nuclear recoil effects are presented separately below using expansion in $\alpha$.
\begin{table}
\caption{Contributions to the ground state HFS of H, $E_\mathrm{hfs}^\mathrm{exp} = 1\,420\,405.751\,766(3)$ kHz \cite{Hmaser1,Hmaser2}, $E_F = 1\,418\,840.091$ kHz.}
\begin{ruledtabular}
\begin{tabular}{lw{1.14}l}
\multicolumn{1}{l}{Term} &
        \multicolumn{1}{c}{Value} &
               \multicolumn{1}{c}{Ref. and comments}
\\
\hline\\[-5pt]
$a_e$  & 0.001\,159\,652 & Ref. \cite{codata22} \\
$\delta^{(2)}$ & -0.000\,016\,340   & Eq. (\ref{118}), Ref. \cite{eides:01}\\
$\delta^{(3)}$ & -0.000\,007\,099   & Eq. (\ref{119}), Ref. \cite{eides:01, heplus}\\
$\delta^{(4)}$ & -0.000\,000\,121   & Eq. (\ref{120}), Ref. \cite{eides:01, yerokhin:08:prl} \\
$\delta_\mathrm{fns}^{(1)}$ &-0.000\, 039\, 835 (113) &  Eq. (\ref{122}), Ref. \cite{Lin2022}\\
$\delta_\mathrm{rec}^{(1)}$ & 0.000\,005\,291(17)     & Eq. (\ref{123}), Ref.  \cite{AA2022a} \\
$\delta_\mathrm{pol}^{(1)}$ & 0.000\,001\,090(310) & Eq. (\ref{124}), Ref. \cite{Ruth2024} \\
$\delta_\mathrm{rel,fns}^{(2+)}$  & -0.000\,000\,029 & Eq. (\ref{126}), this work\\
$\delta^{(2)}_\mathrm{rad,fns}$ &-0.000\,000\,609 &  Eq. (\ref{127}), Ref. \cite{sgk:97}\\
$\delta_\mathrm{rel,rec}^{(2)}$  &  0.000\,000\,575  & Eq. (\ref{128}), this work \\
$\delta_\mathrm{rad,rec}^{(2)}$ & 0.000\,000\,072 & Eq. (\ref{129}), Ref. \cite{rad_rec_hfs} \\
$\delta^{(2)}_\mathrm{\mu vp}$          &  0.000\,000\,072 &Eq. (\ref{130}),  Ref. \cite{sgk:97} \\
$\delta^{(2)}_\mathrm{hvp}$ & 0.000\,000\,061 & Ref. \cite{Lensky2026}, hadronic VP\\
$\delta_\mathrm{weak}$ & 0.000\,000\,058 & Ref. \cite{Eides96}, Weak Force\\ \\
$\delta_\mathrm{theo}$ & 0.001\,102\,838(330) & this work\\
$\delta_\mathrm{exp}$ & 0.001\,103\,480\\
$\delta_\mathrm{exp} - \delta_\mathrm{theo}$ &0.000\,000\,642(330) &  2$\sigma$ discrepancy\\
\end{tabular}
\end{ruledtabular}
\end{table}
$\delta^{(1)}_\mathrm{nuc}$ is the leading nuclear structure correction
coming from the two-photon exchange,
\begin{align}
\delta^{(1)}_\mathrm{nuc} =&\ \delta^{(1)}_\mathrm{fns} + \delta^{(1)}_\mathrm{rec} + \delta^{(1)}_\mathrm{pol}\,. \label{121}
\end{align}
$\delta^{(1)}_\mathrm{fns}$, denoted in the other  literature by $\delta_\mathrm{Z}$, is the so-called Zemach correction \cite{Zemach},
which is given by
\begin{align}
\delta^{(1)}_\mathrm{fns} =&\ -2\,Z\,\alpha\,m\,r_Z\,, \label{122}
\end{align}
where $r_Z$ is the Zemach radius, given by Eq. (\ref{06}). 
Using the modern parametrization of proton form factors, one obtains \cite{Lin2022}  $r_Z = 1.054(3)\,\mathrm{fm}$.
The difference, in comparison to the literature \cite{AA2022a}  is in the presence of a lepton mass instead of the reduced mass.
This difference is absorbed by $\delta^{(1)}_\mathrm{rec}$, which is obtained without expansion in the electron-proton mass ratio \cite{AA2022a}
\begin{align}
\delta^{(1)}_\mathrm{rec} =&\ 5.269^{+0.017}_{-0.004}\,\mathrm{ppm} + 2\,Z\,\alpha\,m\,r_Z\,\frac{m}{m+M}\,, \label{123}
\end{align}
while the sum $\delta^{(1)}_\mathrm{fns} + \delta^{(1)}_\mathrm{rec}$ is by definition a complete  two-photon exchange elastic (Born) amplitude.
Finally, the inelastic contribution includes everything beyond the Born approximation. 
Its evaluation has a long history starting from Ref. \cite{old_pol} in 1967, see Ref. \cite{AA2022b} for a comprehensive review.
This correction dominates uncertainty of  theoretical predictions for the hydrogen HFS, and we take the most recent result from Ref. \cite{Ruth2024},
\begin{align}
\delta^{(1)}_\mathrm{pol} =&\ 1.09(31)\,\mathrm{ppm}\,. \label{124}
\end{align}

The higher order nuclear-structure corrections and the recoil corrections are
much smaller than $\delta^{(1)}_\mathrm{nuc}$. 
They will be calculated by expansion in the electron-proton mass ratio, assuming elastic approximation and that the nuclear charge and
the magnetic moment are the same and given by a dipole parametrization in Eq. (\ref{20}).

$\delta^{(2+)}_\mathrm{rel,fns}$ is the relativistic and higher order nuclear size correction (in the nonrecoil limit).
The leading term in the $Z\,\alpha$ expansion \cite{muD} is
\begin{align}
\delta^{(2)}_\mathrm{rel,fns}=&\ \frac{4}{3}\,(m\,r_C\,Z\,\alpha)^2 \label{125} \\ & \times
\Big[-1+\gamma + \ln(2\,m\,r_{CC}\,Z\,\alpha) + \frac{r_M^2}{4\,r_C^2}\Big]\,, \nonumber
\end{align}
where $r_M$ is the root-mean-square magnetic radius, and $r_{CC}/r_C= 5.274\,565$ for the exponential charge distribution.
 Its numerical value is, however, smaller than the next-order correction. Therefore, instead of using this formula for
 $\delta^{(2)}_\mathrm{rel,fns}$, we calculate finite nuclear size effects by solving numerically the Dirac equation and obtain
\begin{align}
\delta^{(2+)}_\mathrm{rel,fns} = -29\times 10^{-9}\,. \label{126}
\end{align}

The radiative finite nuclear size correction in the nonrecoil limit is \cite{sgk:97}
\begin{align}
 \delta^{(2)}_\mathrm{rad,fns}=&\  -2\,Z\,\alpha\,m\, r_Z
 \,\frac{\alpha}{\pi}\,\biggl( -\frac{5}{4} + \frac{2}{3}\,\ln\frac{\Lambda^2}{m^2} - \frac{634}{315}\biggr)\,, \label{127}
\end{align}
where the first terms comes from the electron self-energy and next two from the vacuum polarization.
Using the relation $\Lambda = 35/(8\,r_Z)$ we obtain the numerical value shown in Table~I.

$\delta^{(2)}_\mathrm{rel,rec}$ is the relativistic recoil correction for a point nucleus given by
\begin{align}
\delta^{(2)}_\mathrm{rel,rec} =&\ \frac{m}{M}\,\frac{(Z\,\alpha)^2}{1+\kappa}\,
\bigg[\frac{65}{18} + \frac{13}{18}\,\kappa + \frac{31}{36}\,\kappa^2 \label{128}
\\ &\hspace*{-7ex}
- \Big(8 + 2\,\kappa - \frac{1}{4}\,\kappa^2\Big)\,\ln2 
- \Big(2 + 2\,\kappa + \frac{7}{4}\,\kappa^2\Big)\,\ln(Z\,\alpha)\bigg]. \notag
\end{align}
It is derived in this work and corrects the previous calculation by Bodwin and Yennie in Ref.  \cite{Bodwin:88}.
Numerically, our result for the hydrogen  $\delta^{(2)}_\mathrm{rel,rec} = 0.575\;\mathrm{ppm}$ does not differ much from that by 
Bodwin and Yennie $(0.464\;\mathrm{ppm})$.

$\delta^{(2)}_\mathrm{rad,rec}$ is the radiative recoil correction, which is a sum of vacuum polarization and self-energy.
Karshenboim \cite{sgk:97} presented only a rough estimation for this correction in hydrogen, and his results are
$\delta^{(2)}_\mathrm{vp,rec} = -0.02$ ppm and $\delta^{(2)}_\mathrm{se,rec} = 0.11(2)$ ppm. 
The result of our calculations  in Ref. \cite{rad_rec_hfs} is 
\begin{align}
\delta_{\text{rad,rec}}^{(2)} =&\ -0.032 + 0.104 = 0.072\;\mathrm{ppm}\,, \label{129}
\end{align}
in agreement with numerical values from Ref. \cite{sgk:97}.

$\delta^{(2)}_\mathrm{\mu vp}$ is the muon vacuum polarization with the finite size nucleus and without subtracting a point nucleus,
\begin{align}
\delta^{(2)}_\mathrm{\mu vp} =&\  \frac{m}{m_\mu}\bigg[
\frac{3}{4}\,   Z \alpha^2
 - 2\,Z \alpha\,m_\mu\,r_Z\,\frac{\alpha}{\pi} \,\bigg(\frac{2}{3}\,\ln\frac{\Lambda^2}{m_\mu^2} -\frac{634}{315} \bigg)
 \nonumber \\ &\   + \Big(\frac{m_\mu}{\Lambda}\Big)^2\bigg]\,. \label{130}
\end{align}
Finally, the hadronic vacuum polarization is  taken from Ref. \cite{Lensky2026}
\begin{align}
\delta_\mathrm{hvp} = 0.061\;\mathrm{ppm}
\end{align}
and weak force correction from Ref. \cite{Eides96}.
The sum of all important contributions in comparison to experimental result is presented in Table~I, 
and we observe $2\,\sigma$ discrepancy with the experimental value.

\section{Specific difference $D_{21}$}
Considering the specific difference $D_{21}$ in HFS
\begin{align}
D_{21}=&\ 8\,E_\mathrm{hfs}(2S)-E_\mathrm{hfs}(1S)\,,
\end{align}
where nuclear structure effects cancel out to a high degree, our result
\begin{align}
\delta D_{21}(\mathrm{new}) =&\ 
(Z\,\alpha)^2\,\frac{m}{M}\,E_F\biggl[2\,\ln 2 -\frac{19}{8} 
\nonumber \\ &\hspace*{-10ex}
+ \biggl(-\frac{7}{2}\,\ln 2 +\frac{73}{32}\biggr)\,\frac{g-1}{g} +\biggl(\frac{7}{8}\,\ln 2 - \frac{145}{128}\biggr)\, g\biggr]
\end{align}
is in disagreement with the previous result from Ref. \cite{D21theory}
\begin{align}
\delta D_{21}(\mathrm{old})  =&\ 
(Z\,\alpha)^2\,\frac{m}{M}\,E_F\biggl[-\frac{9}{8} 
\nonumber \\ &\hspace*{-10ex}
+ \biggl(\frac{\ln 2}{2}-\frac{7}{32}\biggr)\,\frac{g-1}{g} +\biggl(\frac{7}{8}\,\ln 2 - \frac{145}{128}\biggr)\,g \biggr]\,.
\end{align}
The difference
\begin{align}
\Delta_{21} =&\ \delta D_{21}(\mathrm{new}) - \delta D_{21}(\mathrm{old})
\nonumber \\ =&\
(Z\,\alpha)^2\,\frac{m}{M}\,E_F\biggl(\frac{5}{4} - 2\,\ln 2\biggr)\,\,\frac{g-2}{g}
\end{align}
affects theoretical predictions for $D_{21}$ in H and He$^+$, see Table~II.
For H, we used Ref. \cite{yost} for the most recent  and accurate experimental and theoretical values,
and for He$^+$ HFS we took the previous theoretical predictions from Ref. \cite{D21theory}
and experimental results from Refs. \cite{heplus1, heplus2}.
We observe an excellent agreement for He$^+$. While,
in the hydrogen case, the relative experimental uncertainty is large,  so agreement is maintained, 
but a new, more precise measurement of hydrogen $E_\mathrm{hfs}(2S)$ might be in place. 
\begin{table}
\caption{$D_{21}$  with corrected recoil  in H and He$^+$ in Hz.}
\begin{ruledtabular}
\begin{tabular}{lw{5.2} w{10.4}}
\multicolumn{1}{l}{$D_{21}$} &
        \multicolumn{1}{c}{H} &
               \multicolumn{1}{c}{He$^+$}
\\
\hline\\[-5pt]
$D_{21}^\mathrm{the}(\mathrm{old})$                & 48\,954.1(2.3) & -1\,190\,068.(64) \\
$\Delta_{21}$                                                               & -3.6                  & 60.\\
$D_{21}^\mathrm{the}$                                       & 48\,950.5(2.3) & -1\,190\,008.(64) \\
$D_{21}^\mathrm{exp}$                                      & 48\,959.2(6.8) & -1\,189\,979.(71)\\[1ex]
$D_{21}^\mathrm{exp} - D_{21}^\mathrm{the}$ & 8.7(7.2) & 29.(96)\\
\end{tabular}
\end{ruledtabular}
\end{table}

\section{Summary and conclusions}
We have calculated $(Z\,\alpha)^2\,m/M\,E_F$ relativistic recoil correction to the hyperfine splitting in hydrogen-like systems.
Our result is in disagreement with the $n=1$ value obtained by Bodwin and Yennie in Ref. \cite{Bodwin:88}, and also with state dependence from Ref. \cite{D21theory},
but is in agreement  with previous calculations for $g=2$ case.
We observe $2\,\sigma$ discrepancy  of theoretical predictions with the measurement of the hydrogen ground state HFS.
Most probably, it comes from the not well known proton structure effects.
Namely, the current theoretical value for the proton structure contribution $\delta^{(1)}_\mathrm{nuc}(\mathrm{the}) = -33.45(33)$ ppm can be compared
with the proton structure inferred from the comparison of measured HFS with theoretical predictions   
$\delta^{(1)}_\mathrm{nuc}(\mathrm{exp}) = -32.81$ ppm and the difference is $2\,\sigma$.
Planned HFS measurements in $\mu$H \cite{antognini, vacchi} may shed light on the origin of this discrepancy.
It is because one can compare regular H with $\mu$H by considering the difference
\begin{align}
\Delta =&\ \frac{m_p}{m_e}\,\delta^{(1)}_\mathrm{nuc}(\mathrm{H}) - \frac{m_p}{m_\mu}\,\delta^{(1)}_\mathrm{nuc}(\mu\mathrm{H})
\nonumber \\ =&\ \Delta_\mathrm{fns} +  \Delta_\mathrm{rec} +  \Delta_\mathrm{pol} \label{Delta}\,.
\end{align}
This difference $\Delta$ can be calculated much more accurately than individual $\delta^{(1)}_\mathrm{nuc}$,
because the large momentum contributions cancel out.
In fact $\Delta_\mathrm{fns} = 0$, $\Delta_\mathrm{rec}$ using Eq. (\ref{21})
\begin{align} 
\Delta_\mathrm{rec} \approx &\
- \frac{Z\,\alpha}{\pi}\,\frac{2}{g}
 \left( \frac{3\, g^2}{16} - \frac{3\, g}{4} - \frac{9}{4} \right) \ln\left(\frac{m_e}{m_\mu}\right)
\end{align}
can be expressed in terms of kinematic factors, and only the smallest $\Delta_\mathrm{pol}$ requires a separate analysis,
but one may also expect a significant cancellation here. In conclusion, provided QED theory for $\mu$H achieves sufficient accuracy,
one can improve the test of fundamental interactions theory by considering the difference $\Delta$ in Eq. (\ref{Delta}).
The result obtained in this work for the relativistic recoil correction $\delta^{(2)}_\mathrm{rel.rec} = 0.000\,119$ for $\mu$H is indispensable to achieve this goal.

Considering the difference $D_{21} = 8\,E_\mathrm{hfs}(2S) - E_\mathrm{hfs}(1S)$, it is 
very much independent of the proton structure effects. The obtained result for the relativistic recoil correction
improves an agreement for He$^+$ HFS, which probably is the most sensitive test for the existence 
of spin-dependent long-range interactions. However, the obtained result worsens agreement with the hydrogen $D_{21}$ 
This indicates the need for improved measurement of $E_\mathrm{hfs}(2S) $ in hydrogen before drawing  further conclusions.

Finally, it would be worth directly calculating numerically complete  recoil corrections using Eqs. (\ref{25},\ref{26},\ref{27}), 
because one can account for the finite nuclear size effects, which are important for $\mu$H, and also to account for higher powers of $Z\,\alpha$,
which are important in heavier muonic atoms. This project is currently being pursued \cite{Hevler}.

\section*{Acknowledgments}
We gratefully acknowledge interesting discussion with Aldo Antognini and Vadim Lensky.

\appendix

\section{Spin algebra in $d$-dimensions}
\label{AppA}
Apart from the lack of  the Levi-Civita symbol in arbitrary $d$-dimensions, the spin algebra is the same as in $d=3$.
Let us therefore define an antisymmetric tensor $\sigma^{jk}$  for a spin $1/2$ particle, 
\begin{align}
\sigma^{ij} =&\  -\frac{i}{2}\,[\sigma^i\,,\,\sigma^j]\,, \label{A1}
\end{align}
such that in $d=3$
\begin{align}
\sigma^{ij} \stackrel{d\rightarrow 3}{=}&\ \epsilon^{ijk}\,\sigma^k\,. \label{A2}
\end{align}
One can express arbitrary products of $\sigma^i$ in terms of $\delta^{ij}$, $\sigma^i$, and $\sigma^{ij}$, for example
\begin{align}
\sigma^i\,\sigma^j =&\ \delta^{ij} + i\,\sigma^{ij} \label{A3}\\
\sigma^a\,\sigma^b\,\sigma^c\,\sigma^d =&\ \delta^{ad}\,\delta^{bc} - \delta^{ac}\,\delta^{bd} + \delta^{ab}\,\delta^{cd}
\nonumber \\ &\
+ i\,(\delta^{cd}\,\sigma^{ab} + \delta^{bd}\,\sigma^{ca} + \delta^{ac}\,\sigma^{db} 
\nonumber \\ &\
+ \delta^{ad}\,\sigma^{bc} + \delta^{ab}\,\sigma^{cd} + \delta^{bc}\,\sigma^{ad}). \label{A4}
\end{align}
From the last Eq. (\ref{A4}), one obtains
\begin{align}
\{\sigma^{ij}\,,\,\sigma^{kl}\} =&\ 2\,(\delta^{ik}\,\delta^{jl} -\delta^{il}\,\delta^{jk}) \label{A5}\\
[\sigma^{ij}\,,\, \sigma^{kl}] =&\ 2\,i\,\big(\delta^{jl}\,\sigma^{ik} + \delta^{ik}\,\sigma^{jl}+ \delta^{il}\,\sigma^{kj} + \delta^{jk}\,\sigma^{li}\big). \label{A6}
\end{align}
For an arbitrary spin particle we define $S^{ij}$, such that
\begin{align}
S^{ij} \stackrel{d\rightarrow 3}{=}&\ \epsilon^{ijk}\,S^k\,,
\end{align}
where $\vec S$ is the spin operator and  $\vec S = \vec\sigma/2$ for $S=1/2$.
The commutator of general spin operators is similar to Eq. (\ref{A6})
\begin{align}
[S^{ij}\,, S^{kl}] =&\ i\,\big(\delta^{jl}\,S^{ik} + \delta^{ik}\,S^{jl}+ \delta^{il}\,S^{kj} + \delta^{jk}\,S^{li}\big),
\end{align}
and the angular average in $d$-dimensions leads to
\begin{align}
\langle S^{ij}\,\sigma^{kl}\rangle =&\ S\cdot\sigma\,\frac{\delta^{ik}\,\delta^{jl} -\delta^{il}\,\delta^{jk}}{d\,(d-1)}\,, \\
\langle S^{ij}\,\sigma^{il}\rangle  =&\ S\cdot\sigma\,\frac{\delta^{jl}}{d}\,, 
\end{align}
where
\begin{align}
S\cdot\sigma = S^{ab}\,\sigma^{ab} \stackrel{d=3}{=} 4\,\vec S\cdot\vec s\,.
\end{align}

\section{Bound-state matrix elements in $d$-dimensions}
\label{AppB}
Let us define
\begin{align}
{\cal V}_1(r) = \int \frac{d^dk}{(2\,\pi)^d}\,\frac{4\,\pi}{k^2}\,e^{i\,\vec k\cdot \vec r} =&\ \frac{C_1}{r^{1-2\,\epsilon}} \equiv \frac{1}{r_\epsilon}\,,
\\
{\cal V}_2(r) = \int \frac{d^dk}{(2\,\pi)^d}\,\frac{4\,\pi}{k^4}\,e^{i\,\vec k\cdot \vec r} =&\ C_2\,r^{1+2\,\epsilon} \equiv -\frac{1}{2}\,r_\epsilon\,,
\end{align}
where
\begin{align}
C_1 =&\ \pi^{\epsilon-1/2}\,\Gamma(1/2 -\epsilon)\,,
\\
C_2 =&\  \frac{1}{4}\,\pi^{\epsilon-1/2}\,\Gamma(-1/2 -\epsilon)\,.
\end{align}
Many integrals can be expressed in terms of ${\cal V}_1$ and ${\cal V}_2$, for example
\begin{align}
 \frac{1}{2}\,\biggl[\frac{\delta^{ij}}{r} + \frac{r^i\,r^j}{r^3}\biggr]_\epsilon \equiv&\
\int \frac{d^dk}{(2\,\pi)^d}\,\frac{4\,\pi}{k^2}\,\biggl(\delta^{ij}-\frac{k^i\,k^j}{k^2}\biggr)\,e^{i\,\vec k\cdot \vec r}
\nonumber \\ =&\ 
\delta^{ij}\,{\cal V}_1 + \partial^i\,\partial^j\,{\cal V}_2 
\end{align}
and
\begin{align}
4\,\pi\,\delta_\perp^{ij} =&\  \int \frac{d^d k}{(2\,\pi)^d}\,4\,\pi\, \biggl(\delta^{ij}-\frac{k^i\,k^j}{k^2}\biggr)\,e^{i\,\vec k\cdot\vec r}
\nonumber \\  =&\ 
\frac{(d-1)}{d} \delta^{ij} 4\,\pi\,\delta^d(r) +
\biggl(\partial^i\partial^j-\frac{\delta^{ij}}{d}\,\partial^2\biggr)\,{\cal V}_1
\nonumber\\ =&\
 \biggl[\frac{2}{3}\,\delta^{ij}\,4\,\pi\,\delta^3(r) + 
\frac{1}{r^5}\,(3\,r^i\,r^j-\delta^{ij}\,r^2)\biggr]_\epsilon\,.
\end{align}
The following matrix elements frequently appear in calculations
\begin{align}
\bigg\langle \frac{1}{r^4_\epsilon}\bigg\rangle \equiv &\ \bigg\langle\bigg[ \vec \nabla \frac{1}{r_\epsilon}\bigg]^2\bigg\rangle
= \bigg\langle\frac{1}{r^4}\bigg\rangle + \langle \pi\,\delta^d(r)\rangle\bigg(-\frac{2}{\epsilon} + 8\bigg),
\label{B1}\\
\bigg\langle\frac{1}{r^3_\epsilon}\bigg\rangle \equiv &\ \bigg\langle\bigg[\frac{1}{r_\epsilon}\bigg]^3\bigg\rangle
= \bigg\langle\frac{1}{r^3}\bigg\rangle + \langle \pi\,\delta^d(r)\rangle\bigg(\frac{1}{\epsilon} + 2\bigg)\,,
\label{B2}
\end{align}
where for $l=0$ states
\begin{align}
 \bigg\langle\frac{1}{r^3}\bigg\rangle  =&\ \lim_{a\rightarrow0}\int_a^\infty \frac{d r}{r} \, f(r) + f(0)\,(\gamma+\ln a)
 \nonumber \\ =&\ 
\frac{4}{n^3}\bigg(\frac12 - \frac{1}{2\,n} - \gamma - \Psi(n) + \ln\frac{n}{2}  \bigg),
 \\
 \bigg\langle\frac{1}{r^4}\bigg\rangle  =&\ \lim_{a\rightarrow0}\int_a^\infty \frac{d r}{r^2} \, f(r) -\frac{f(0)}{a}+ f'(0)\,(\gamma+\ln a)
 \nonumber \\ =&\  
\frac{8}{n^3}\bigg(-\frac53 + \frac{1}{2\,n} + \frac{1}{6\,n^2} + \gamma + \Psi(n) - \ln\frac{n}{2} \bigg)\,,
\end{align}
with
\begin{align}
 f(r) = &\ \int d\Omega\, \phi^*(\vec r)\,\phi(\vec r)\,.
\end{align}
Further identities involving singular matrix elements
\begin{align}
\bigl\langle 4\,\pi\,\delta_\perp^{ij}\,p^i\,p^j\bigr\rangle =&\  -\langle\pi\,\nabla^2\delta^d(r)\rangle - \bigg\langle \frac{1}{r^4_\epsilon}\bigg\rangle\,, \\
\biggl\langle \delta^d(r)\,\frac{1}{r_\epsilon} \biggr\rangle =&\ 0\,.
\end{align}
The regular matrix elements for $l=0$ states are
\begin{align}
E = &\ - \frac{1}{2\,n^2}\,, \label{B6}
\\
\bigg\langle \frac{1}{r}\bigg\rangle = &\ \frac{1}{n^2}, \label{B7}
\\
\bigg\langle \frac{1}{r^2}\bigg\rangle = &\ \frac{2}{n^3}, \label{B8}
\\
\langle \pi\,\delta^3(r)\rangle = &\ \frac{1}{n^3},  \label{B11}
\\
\big\langle \pi\,\nabla^2 \delta^d(r) \big\rangle =&\ \frac{2}{n^5}\,,  \label{B12}
\\
\big\langle \big\{ p^i , \big\{ p^i , \pi\,\delta^d(r)\big\}\!\big\} \big\rangle =&\ -\frac{2}{n^5}\,,  \label{B13}
\\
\bigg\langle p^i\,\bigg(\frac{\delta^{ij}}{r} + \frac{r^i r^j}{r^3}\bigg)p^j\bigg\rangle = &\ \frac{1}{n^3}\bigg(-\frac{2}{n}+4\bigg)\,,  \label{B14}
\\
\bigg\langle p^i\,\bigg(\frac{\delta^{ij}}{r^2} + \frac{r^i r^j}{r^4}\bigg)p^j\bigg\rangle = &\ \frac{1}{n^3}\bigg(-\frac{4}{3\,n^2}+\frac{16}{3}\bigg)\,. \label{B15}
\end{align}

Second-order matrix elements for $l=0$ states are
\begin{widetext}
\begin{align}
\bigg\langle \pi\,\delta^d(r)\frac{1}{(E-H)'} \frac{p^4}{8}  \bigg\rangle =&\ \frac{1}{n^3}\bigg(-\frac32 - \frac{1}{n} + \frac{5}{4\,n^2} + \gamma + \Psi(n) - \ln\frac{n}{2} \bigg)
- \frac{1}{4\,\epsilon}\langle\pi\,\delta^d(r)\rangle
 \,, \label{B3}\\
\bigg\langle \pi\,\delta^d(r) \frac{1}{(E-H)'}  \pi\,\delta^d(r) \bigg\rangle =&\ \frac{1}{n^3}\bigg(- \frac12  - \frac{1}{n} + \gamma + \Psi(n) - \ln\frac{n}{2}  \bigg) - \frac{1}{4\,\epsilon}\langle\pi\,\delta^d(r)\rangle
\,, \label{B4}\\
\bigg \langle \pi\,\delta^d(r)\frac{1}{(E-H)'} \frac{1}{2}\,p^i\,\bigg[\frac{\delta^{ij}}{r} + \frac{r^i r^j}{r^3}\bigg]_\epsilon\,p^j  \bigg \rangle =&\
\frac{2}{n^3}\bigg(-\frac94 - \frac{1}{n} + \frac{5}{4\,n^2} + \gamma + \Psi(n) - \ln\frac{n}{2} \bigg) - \frac{1}{2\,\epsilon}\langle\pi\,\delta^d(r)\rangle\,,
\\
\biggl\langle \bigg[\frac{\delta^{ij}}{r^3}-3\frac{r^i r^j}{r^5}\bigg]_\varepsilon\,\frac{1}{E-H}
\bigg[\frac{\delta^{ij}}{r^3}-3\frac{r^i r^j}{r^5}\bigg]_\varepsilon \biggr\rangle
=&\ 
 -\frac{40}{3\,n^3}\,\bigg(-\frac{8}{15} + \frac{1}{2\,n} + \frac{3}{20\,n^2} + \gamma + \Psi(n) - \ln\frac{n}{2} \bigg)
 + \frac{10}{3\,\epsilon}\,\langle \pi\,\delta^d(r)\rangle\,. 
\end{align}

\section{Integration of the high-energy part}
\label{AppD}
The integration over $\vec k_1$ and $\vec k_2$ is performed using the following formula
 \begin{align}
&\ \int\frac{d^d k_1}{(2\,\pi)^d}\, \int\frac{d^d k_2}{(2\,\pi)^d}\,
\frac{4\,\pi}{(k_1^2)^{n_1}}\, \frac{4\,\pi}{(k_2^2 + m_2^2)^{n_2}}\,\frac{4\,\pi}{(k_3^2 + m_3^2)^{n_3}}
\nonumber \\ =&\
  \frac{m_3^{2\,(d-n_1-n_2-n_3)}}{(4\,\pi)^{d-3}}\,
  \frac{\Gamma(d/2-n_1)\, \Gamma(n_1+n_2-d/2)\, \Gamma(n_1+n_3-d/2)\, \Gamma(n_1+n_2+n_3-d)}
       {\Gamma(2\,n_1+n_2+n_3-d)\, \Gamma(n_2)\, \Gamma(n_3)\, \Gamma(d/2)}\nonumber \\&\
\times       _2\!F_1(n_1+n_2+n_3-d,n_1+n_2-d/2,2\,n_1+n_2+n_3-d,1- m_2^2/m_3^2)\,. 
\end{align}
\end{widetext}
After $\vec k$-integration, the integrand $f$ is a real function of $\omega$,
which is analytic for $\Re(\omega)>0$ and for  $\Re(\omega)<0$. 
The Feynman integration contour is
\begin{align}
 \int_0^\infty d\,\omega\,\big[ f(\omega+i\,\epsilon) - f(\omega-i\,\epsilon)\big]\,.
\end{align}
We assume that the square root has a branch cut on a real negative axis, so
\begin{align}
\sqrt{z} =&\ \sqrt{|z|\,e^{i\,\phi}} = \sqrt{|z|}\,e^{i\,\phi/2}
\end{align}
for $-\pi<\phi<\pi$, and for an arbitrary real $n$
\begin{align}
z^n =&\ |z|^n\,e^{i\,\phi\,n}\,.
\end{align}
Our integrand contains $(-\omega^2)^n$, which shall be interpreted for $\Re(\omega)>0$ and $\Im(\omega)>0$ as
\begin{align}
\big(-\omega^2\big)^n =&\ \big(-(\omega+i\,\epsilon)^2\big)^n  = \omega^{2\,n}\,e^{-i\,\pi\,n}\,,
\end{align}
therefore
\begin{align}
\big(-(\omega+i\,\epsilon)^2\big)^n  - \big(-(\omega-i\,\epsilon)^2\big)^n =&\ -2\,i\,\sin(\pi\,n)\,\omega^{2\,n}\,.
\end{align}
Our integrand contains also $\big(-2\,\omega-\omega^2\big)^n$, which shall be interpreted for $\Re(\omega)>0$  as 
\begin{align}
\big(-2\,\omega-\omega^2\big)^n =&\ \biggl(1+\frac{2}{\omega}\biggr)^n\,\big(-\omega^2\big)^n\,.
\end{align}
So, the $\omega$ integration with a regular $f$ is
\begin{align}
&\ i\,\bigg( \int^{0-i\,\epsilon}_{\infty-i\,\epsilon} + \int_{0+i\,\epsilon}^{\infty+i\,\epsilon} \bigg) d\,\omega\,(-\omega^2)^n\,f(\omega)
\nonumber \\ 
=&\ 2\,\sin(\pi\,n)\,\int_0^\infty d\,\omega\,\omega^{2\,n}\,f(\omega)\,.
\end{align}
After changing the integration variable  to $z = 2/\omega$
\begin{align}
\int_0^\infty d\,\omega\,f(\omega) =&\ \int_0^\infty dz\,\frac{2}{z^2}\,f\bigg(\frac{2}{z}\bigg)
\end{align}
the last $z$-integral can be performed using
\begin{align}
& \int_0^\infty {}_2F_1(a,b,c; -t)\,t^{-s-1}\,dt \\
=&\ \Gamma(-s)\,\frac{\Gamma(a+s)}{\Gamma(a)}\,\frac{\Gamma(b+s)}{\Gamma(b)}\,\,\frac{\Gamma(c)}{\Gamma(c+s)}
\end{align}
and
\begin{align}
\int_0^\infty \frac{t^{a-1}}{(1+t)^b}\,dt =&\ \frac{\Gamma(a)\,\Gamma(b-a)}{\Gamma(b)}\,.
\end{align}

\section{Atomic units in $d$-dimensions}
\label{AppC}
The dimensional regularization is performed in natural units, while
the elimination of singularities will be performed in atomic units,
which in $d$-dimensions become a little more complicated.
The nonrelativistic Hamiltonian in natural units is
\begin{equation}
H = \frac{\vec{p}^{\,2}}{2\,m} -Z\,\alpha\,\frac{C_1}{r^{1-2\,\epsilon}}\,. \label{C1}
\end{equation}
Changing coordinates to atomic units
\begin{equation}
\vec r \rightarrow (m\,Z\,\alpha)^{-1/(1+2\,\epsilon)}\,\vec r, \label{C2}
\end{equation}
$H$ can be written as
\begin{align}
H = m^{(1-2\,\epsilon)/(1+2\epsilon)}\,(Z\,\alpha)^{2/(1+2\,\epsilon)}\,
\biggl[\frac{\vec{p}^{\,2}}{2} -\frac{C_1}{r^{1-2\,\epsilon}}\biggr]. \label{C3}
\end{align}
If one pulls out the factor
$m^{(1-2\,\epsilon)/(1+2\epsilon)}\,(Z\,\alpha)^{2/(1+2\,\epsilon)}$
from $H$, then one will obtain the nonrelativistic Hamiltonian in atomic units.
Similarly for $H^{(6)}$, the common factor in atomic units 
\begin{align}
m^{(1-10\,\epsilon)/(1+2\epsilon)}\,(Z\,\alpha)^{6/(1+2\,\epsilon)} = m\,(Z\,\alpha)^6\,\eta \label{C4}
\end{align}
is pulled out. Such a factor will also be pulled out
from $E_H$, which will lead to the appearance of the logarithmic term, namely
with $m=1$
\begin{align}
\frac{(Z\,\alpha)^3\,}{\epsilon}\,\delta^d(r) \rightarrow &\  
(Z\,\alpha)^{-\frac{6}{1+2\,\epsilon}}\,\frac{(Z\,\alpha)^3}{\epsilon}\,(Z\,\alpha)^{\frac{3-2\,\epsilon}{1+2\,\epsilon}}\,\delta^d(r)
 \nonumber \\ 
\nonumber \\ =&\
\biggl(\frac{1}{\epsilon} + 4\,\ln(Z\,\alpha)\biggr)\,\delta^d(r)\,. \label{C5}
\end{align}

\end{document}